\documentclass[lettersize,journal,twocolumn]{IEEEtran}
\usepackage{amssymb,amsmath}
\usepackage{cite}
\usepackage{graphicx}
\usepackage{psfrag}
\usepackage{url}
\usepackage[T1]{fontenc}
\usepackage[utf8]{inputenc}
\usepackage[absolute,overlay]{textpos}
\usepackage{gensymb}
\usepackage{cases}
\usepackage{cuted}
\usepackage[font=footnotesize]{caption}
\usepackage[font=footnotesize]{subcaption}
\usepackage{float}
\usepackage[linesnumbered,ruled,lined]{algorithm2e}
\SetKwInOut{Training}{Train}
\SetKwInOut{Testing}{Test}
\usepackage{pgf}

\usepackage[nocomma]{optidef}

\usepackage{amsthm}

\usepackage{booktabs}
\usepackage{color,soul}

\usepackage{textcomp}
\usepackage{xcolor}
\def\BibTeX{{\rm B\kern-.05em{\sc i\kern-.025em b}\kern-.08em
    T\kern-.1667em\lower.7ex\hbox{E}\kern-.125emX}}

\usepackage{tabularx}
\usepackage{makecell}
\usepackage{multirow}

\usepackage{graphicx}
\usepackage{caption}
\usepackage{subcaption}

\usepackage{pgf}
\usepackage[nocomma]{optidef}

\usepackage{amsthm}

\usepackage{booktabs}
\usepackage{color,soul}

\begin{document}

%\title{Minimization of the IoT Devices Peak Energy Consumption: A AUV Assisted Approach}
\title{Multi-AUV Trajectory Learning for Sustainable Underwater IoT with Acoustic Energy Transfer}
\author{
	\IEEEauthorblockN{Mohamed Afouene Melki, Mohammad Shehab, and Mohamed-Slim Alouini \\
	}
	\thanks{This work is supported by an ERIF/OSSARI Grant.}
	\thanks{The authors are with CEMSE Division, King Abdullah University of Science and Technology (KAUST), Thuwal 23955-6900, Saudi Arabia (emails: mohamed.melki@kaust.edu.sa, mohammad.shehab@kaust.edu.sa, slim.alouini@kaust.edu.sa).}
}
\maketitle

%intro - appendix - abstract - rest of paper - 
%afouene: abstract - conclusion

\begin{abstract}

The Internet of Underwater Things (IoUT) supports ocean sensing and offshore monitoring but requires coordinated mobility and energy-aware communication to sustain long-term operation. This letter proposes a multi-AUV framework that jointly addresses trajectory control and acoustic communication for sustainable IoUT operation. The problem is formulated as a Markov decision process that integrates continuous AUV kinematics, propulsion-aware energy consumption, acoustic energy transfer feasibility, and Age of Information (AoI) regulation. A centralized deep reinforcement learning policy based on Proximal Policy Optimization (PPO) is developed to coordinate multiple AUVs under docking and safety constraints. The proposed approach is evaluated against structured heuristic baselines and demonstrates significant reductions in average AoI while improving fairness and data collection efficiency. Results show that cooperative multi-AUV control provides scalable performance gains as the network size increases.

\end{abstract}
%

%The Internet of Underwater Things (IoUT) enables persistent ocean monitoring and infrastructure maintenance but remains constrained by limited device energy and harsh channel conditions. This letter presents a multi-AUV framework for simultaneous acoustic energy transfer (AET) and information uplink, targeting enhanced data freshness and sustainability. Using the Age of Information (AoI) as a performance metric, a deep-reinforcement-learning policy based on Proximal Policy Optimization (PPO) reduces AoI by up to 55%, improves data delivery by 40–45%, and maintains a fairness index above 0.9 compared with conventional Round Robin and Random Walk benchmarks.

\vspace{-3mm}
\section{Introduction}

Underwater data collection is vital for applications including environmental monitoring, marine infrastructure inspection, and disaster management. Despite its importance, underwater communication remains challenging due to limited bandwidth, severe signal attenuation, and high latency in acoustic channels~\cite{theocharidis2025underwater}. Advances in communication technologies, such as acoustic and optical solutions and adaptive modulation schemes, are set to address these limitations~\cite{theocharidis2025underwater,busacca2024adaptive}.

To ensure reliable and sustainable data acquisition, AUVs have been increasingly deployed as mobile relays and data collectors in underwater sensor networks. Various studies have shown that optimizing AUV trajectories is essential for improving both energy efficiency and communication reliability in such environments~\cite{9790816,10909654}. Furthermore, reinforcement learning techniques have been leveraged to enable adaptive control and energy-aware decision-making, demonstrating significant gains in the long-term throughput and sustainability of underwater systems~\cite{8882231}. These findings highlight the growing potential of learning-driven trajectory design and energy management to support efficient and robust underwater IoT operations.

Coordination among multiple AUVs introduces additional complexity, particularly for safety and collision avoidance. Techniques such as event-triggered control mechanisms and safe trajectory planning under communication delays have been developed to address these multi-agent challenges~\cite{sun2025intermittent}. Additionally, collision-avoidance strategies tailored for underwater sensor clustering have shown significant throughput and stability improvements~\cite{xue2025collision}. Beyond data freshness, underwater IoUT nodes are fundamentally constrained by limited onboard energy. Battery replacement or periodic retrieval of seabed devices is costly, risky, and often infeasible at depth or scale. To address this, acoustic energy transfer (AET) has emerged as a promising approach that uses acoustic waves to wirelessly deliver power through the water medium, enabling battery-less and maintenance-free operation of underwater sensor networks~\cite{9217956}.

Our previous work~\cite{10909654} addressed single-AUV trajectory learning for underwater acoustic energy transfer and data collection. In this letter, we extend that framework as follows:
\begin{itemize}
\item We extend the problem to a coordinated multi-AUV setting and integrate a realistic mobility-aware energy model within the learning framework, capturing propulsion-related energy expenditure, drag effects, battery evolution, and collision-aware navigation.
\item We adopt a continuous 2D motion control formulation that models heading and speed evolution, enabling more physically consistent trajectory generation compared to discrete directional movements.
\item We evaluate the proposed scheme in terms of AoI minimization, collected data, and Jain fairness across IoUT nodes, and compare its performance against a greedy baseline strategy.
\end{itemize}

\section{System Model}\label{sec:system}

%------------------------------------------------------------------------
\subsection{Layout}
\label{sec:network}

We consider an IoUT scenario comprising \(N\) autonomous underwater vehicles (AUVs) and \(K\) static sensor nodes deployed over a bounded two-dimensional operational area. Each sensor node \(k\) is located at a fixed position \( \mathbf{c}_k = (x_k, y_k) \), while the position of the \(n^{\text{th}}\) AUV at time \(t\) is denoted by \( \boldsymbol{\ell}_{\mathrm{auv}_n}(t) = (x_{\mathrm{auv}_n}(t), y_{\mathrm{auv}_n}(t)) \). 

Each node monitors environmental parameters such as temperature, pH, or dissolved oxygen and communicates with the AUVs through acoustic modems. During navigation, each AUV can (i) perform acoustic energy transfer (AET) to replenish the energy of selected sensor nodes and (ii) collect sensed data via acoustic uplink communication.

\subsection{Channel Model}
In underwater acoustic communication, the dB received level (RL) at an IoUT device located at a distance $d$ from the acoustic source (AUV) can be computed using the sonar equation \cite{6266681} as $RL = SL - AL - NL$, where $SL$ is the acoustic source level, $AL$ represents the total attenuation level, and $NL$ is the ambient noise level. SL is given by  \vspace{-1mm}
\begin{equation}
SL = 170.8 + 10\log_{10}(P_{\text{elec}}) + 10\log_{10}(\eta) + DI,
\end{equation}
where $P_{\text{elec}}$ denotes the source electrical input power, $\eta$ is the electro-acoustic power conversion efficiency , and $DI$ represents the directivity index.

Considering deep water and assuming quasi-static fading, the total attenuation level (AL) is expressed as
\begin{equation}
AL = k_{s}\cdot 10\log_{10}(d) + d\cdot \alpha(f),
\end{equation}

where $k_s$ is the spreading factor and $\alpha(f)$ is the frequency-dependent absorption coefficient modeled by Thorps formula  \cite{9217956}  \vspace{-1mm}
\begin{equation}
\alpha(f) = 0.11\frac{f^2}{f^2 + 1} + 44\frac{f^2}{f^2 + 4100} + 2.75 \times 10^{-4}f^2 + 0.003.
\end{equation}

\vspace{-3mm}
\section{Problem formulation}\label{sec:problem}
\subsection{Energy Harvesting and Information Transmission}

\subsubsection{Acoustic Energy Transfer (AET)}
The available harvestable acoustic power at the IoUT node can be determined as $P_{\text{harv}} = \eta_{\text{harv}} \cdot \frac{10^{\frac{RL + RVS}{10}}}{4R_p} $, where $\eta_{\text{harv}}$ is the harvesting efficiency, $RVS$ is the receiving voltage sensitivity, $R_p$ is the load resistance for impedance matching. The harvested energy over duration $\tau_{\text{charging}}$ is %\footnote{Herein, due to the small Doppler spread and slow channel variations of underwater environments, we assume that the coherence time is long enough to allow for acquiring the Rician channel state information at the transmitter and perform pre-equalization for the channel coefficient.} \vspace{-1mm}
\begin{equation}
E_{\text{harv}} = P_{\text{harv}} \cdot \tau_{\text{charging}}.
\end{equation}

\subsubsection{Information Uplink}
For information uplink transmission from IoUT nodes to the AUV, the required signal-to-noise ratio (SNR) is determined as \vspace{-1mm}
\begin{equation}
\gamma_{\text{req,dB}} \;=\; 10\log_{10}\!\bigl( \; {2^{\mathcal{S}/B}-1} \bigr).
\end{equation} 
Using the passive sonar equation, the required source level $SL_{\text{req}}$ to achieve this SNR is calculated by adding transmission loss (TL) and noise level within the bandwidth ($NL_{\text{band}}$)
\begin{equation}
SL_{\text{req}} = \gamma_{\text{req, dB}} + TL + NL_{\text{band}}.
\end{equation} 
Finally, the transmit power required by node $k$ is computed as
\begin{equation}
P_{\text{trans},k} = 10^{\frac{SL_{\text{req}} - 170.8 - 10\log_{10}\eta_{\text{tx}} - DI_{\text{tx}}}{10}},
\end{equation}
where $\eta_{\text{tx}}$ is the electro-acoustic conversion efficiency of the transmitter, and $DI_{\text{tx}}$ is its directivity index. Then, the energy required for transmission over duration $\tau_{\text{data}}$ is
\begin{equation}
E_{\text{req},k} = P_{\text{trans},k} \cdot \tau_{\text{data}}.
\end{equation}
\subsubsection{Age of Information}
AoI measures the freshness of information as the time elapsed since the last successfully received update. For each sensor node $k \in \{1,\dots,K\}$, we maintain an AoI metric $A_k(t)$ and a service counter $C_k(t)$. To ensure transmission reliability, the AoI resets only after $K_{\text{reset}}$ consecutive successful deliveries. The service counter evolves as
\begin{equation}
\label{AOI_SERVICE}
	C_{k}(t+1) =
	\begin{cases}
		C_{k}(t) + 1, & \text{if delivery occurs at } t, \\
		C_{k}(t), & \text{otherwise},
	\end{cases}
\end{equation}
and the AoI update rule is
\begin{equation}
\label{AOI_CALC}
	A_{k}(t+1) =
	\begin{cases}
		1, & \text{if } C_{k}(t+1) = K_{\text{reset}}, \\
		\min(A_{k}(t) + 1, A_{\text{max}}), & \text{otherwise}.
	\end{cases}
\end{equation}
Upon reset, $C_k(t+1)$ is set to zero.
. %This mechanism prevents premature AoI resets from single isolated transmissions and ensures that information freshness is achieved only through sustained service delivery. The AoI increments by one time unit per step until the service threshold is met.
\subsection{AUV Kinematics} %and Energy Consumption Model}
We consider a discrete-time kinematic model for the AUV operating in a bounded plane. The position of the AUV at time slot $t$ is denoted by $\boldsymbol{\ell}_{\mathrm{auv}}(t) $, and its motion is characterized by a heading angle $\theta(t)$ and a scalar velocity $v(t)$.
Herein, the kinematic evolution is governed by
\begin{align}
\theta(t{+}1) &= \theta(t) + \Delta \theta(t), \\
v(t{+}1) &= v(t) + \Delta v(t),
\end{align}
where $\Delta \theta(t)$ and $\Delta v(t)$ are the control updates for the heading angle and velocity, respectively. The resulting position update follows \vspace{-1mm}
\begin{equation}
\boldsymbol{\ell}_{\mathrm{auv}}(t{+}1)
=
\boldsymbol{\ell}_{\mathrm{auv}}(t)
+
v(t) \Delta t
\begin{bmatrix}
\cos \theta(t) \\
\sin \theta(t)
\end{bmatrix},
\end{equation}
where $\Delta t$ denotes the slot duration. The distance traveled by the AUV during time slot $t$ is therefore
\begin{equation}
d_{\mathrm{auv}}(t)
=
\left\|
\boldsymbol{\ell}_{\mathrm{auv}}(t{+}1)
-
\boldsymbol{\ell}_{\mathrm{auv}}(t)
\right\|.
\end{equation}

The AUV energy consumption is dominated by propulsion and hotel loads. Following \cite{bellingham2009platforms}, the instantaneous propulsion power is modeled as a cubic function of the velocity, yielding
\begin{equation}
P_{\mathrm{prop}}(t)
=
\frac{\rho C_d S}{2\eta_{\mathrm{prop}}} v^3(t) + H,
\end{equation}
where $\rho$ is the seawater density, $C_d$ is the drag coefficient, $S$ is the reference cross-sectional area, $\eta_{\mathrm{prop}}$ is the propulsion efficiency, and $H$ denotes the constant hotel power consumption. Accordingly, the propulsion energy consumed by the AUV during time slot $t$ is given by
\begin{equation}
E_{\mathrm{prop}}(t)
=
\left(
\frac{\rho C_d S}{2\eta_{\mathrm{prop}}} v^3(t) + H
\right)
\frac{d_{\mathrm{auv}}(t)}{v(t)},
\end{equation}
which explicitly couples the AUV energy expenditure to its motion dynamics and trajectory. This formulation enables realistic modeling of the trade-off between mobility, energy consumption, and communication performance.
\begin{figure}[t!]
	\centering
	\includegraphics[width=0.96\columnwidth]{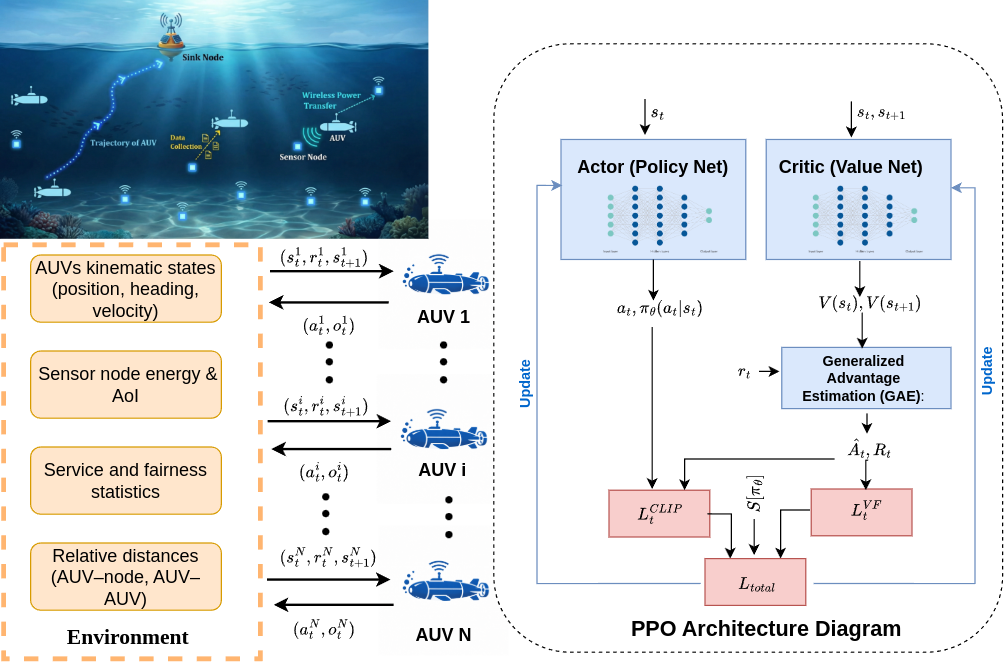} \vspace{0mm}
	\caption{PPO-based interaction between AUVs and the environment} 
	\label{interaction auv env} \vspace{-3mm}
\end{figure}
\vspace{-1.5mm}
\subsection{Problem Formulation}

We consider a finite-horizon AUV-assisted data collection problem involving
$N$ AUVs and $K$ sensor nodes.
The mission duration is not fixed \emph{a priori} and is upper bounded by a
maximum allowable horizon $T_{\max}$.
Let $T$ denote the actual mission completion time, satisfying
$1 \le T \le T_{\max}$.
At each time slot $t$, each AUV jointly determines its motion control inputs
and communication decisions.

The optimization variables include the trajectories of all AUVs,
their incremental motion controls, and the wireless energy transfer (WET)
and data collection decisions.
The problem is formulated as
\begin{equation}
\textbf{P1}:\!
\min_{\substack{
\boldsymbol{\ell}_{\mathrm{auv}_n}(t),\,\Delta\theta_n(t),\,\\\Delta v_n(t),
\mathbf{w}_n(t),\,\mathbf{i}_n(t),\\
t=1,\ldots,T;\; n=1,\ldots,N
}} \!
\frac{1}{T}
\sum_{t=1}^{T} \!
\Bigg(\!
\frac{1}{K}\sum_{k=1}^{K} A_k(t)
+
\lambda_{\mathrm{f}}\big(1-\mathcal{J}(t)\!\big)
\Bigg),
\label{eq:P1_multi_obj}
\end{equation}

\noindent\textbf{s.t.}
\begin{subequations}
\begin{align}
& \boldsymbol{\ell}_{\mathrm{auv}_n}(t) \in \mathcal{X},\;
\boldsymbol{\ell}_{\mathrm{auv}_n}(T) \in \mathcal{G},
&& \forall n,t,
\tag{12a}\label{P1:b} \\[3pt]
& 0 \le v_n(t) \le v_{\max},
&& \forall n,t,
\tag{12b}\label{P1:c} \\[3pt]
& |\Delta\theta_n(t)| \le \Delta\theta_{\max},\;
|\Delta v_n(t)| \le \Delta v_{\max},
&& \forall n,t,
\tag{12c}\label{P1:d} \\[3pt]
& \sum_{k=1}^{K} i_{n,k}(t) = 1,\;
i_{n,k}(t)\in\{0,1\},
&& \forall n,t,
\tag{12d}\label{P1:e} \\[3pt]
& \sum_{k=1}^{K} w_{n,k}(t) = 1,\;
w_{n,k}(t)\in\{0,1\},
&& \forall n,t,
\tag{12e}\label{P1:f} \\[3pt]
& i_{n,k}(t)\,E_{\mathrm{req},k}(t) \le e_k(t),
&& \forall n,k,t,
\tag{12f}\label{P1:g} \\[3pt]
& A_k(t) \le A_{\max},
&& \forall k,t.
\tag{12g}\label{P1:h}
\end{align}
\end{subequations}

In this formulation, $A_k(t)$ denotes the Age of Information (AoI) of node $k$
at time $t$, The fairness term uses Jain's index computed from the empirical service counts
$\mathbf{o}(t)=[o_1(t),\ldots,o_K(t)]$
\vspace{-4mm}
\begin{equation}
\label{eq:jain_env}
\mathcal{J}(t)=\frac{\big(\sum_{k=1}^{K} o_k(t)\big)^2}{K\sum_{k=1}^{K} o_k^2(t)}.
\end{equation}
Constraints in ~\eqref{P1:b} restricts the AUV trajectories to the bounded operational
region $\mathcal{X}$ and enforces terminal goal condition each AUV .
Constraints~\eqref{P1:c}--\eqref{P1:d} impose feasibility on the AUV motion by
bounding the speed and incremental control inputs.
Constraints~\eqref{P1:e} and~\eqref{P1:f} ensure that each AUV selects exactly
one node for uplink data transmission and one node for wireless energy transfer
per time slot, consistent with the FDD operation.
Constraint~\eqref{P1:g} enforces node energy causality, while
constraint~\eqref{P1:h} caps the AoI to prevent unbounded growth.

The above formulation constitutes a Mixed-Integer Nonlinear Program (MINLP) with time coupling and multi-agent coupling.Such problems are generally NP-hard and computationally intractable for large-scale settings using classical optimization techniques. \vspace{-1mm}
\subsection{Greedy AoI-Based Baseline}
The greedy baseline follows a deterministic time-aware navigation strategy that steers the AUV toward the docking point while adapting its speed based on the remaining mission time and spatial constraints, ensuring arrival exactly at the end of the mission horizon. AET is performed greedily to the closest node, while data collection follows a round-robin schedule independent of mobility decisions. This baseline captures intuitive heuristic behavior without leveraging learning or long-term optimization.

\section{The Proposed DRL Solution}\label{sec:solution}

\subsection{Markov Decision Process Formulation}
\label{sec:mdp}

\label{sec:mdp}

We formulate the multi-AUV-assisted data collection and acoustic energy transfer problem
as a finite horizon Markov Decision Process (MDP) defined by the tuple
$\langle \mathcal{S}, \mathcal{A}, R, P \rangle$ over an episode of at most $T_{\max}$
time slots. At each time slot $t$, the system observes the global state
$s(t)\in\mathcal{S}$, selects a joint action $a(t)\in\mathcal{A}$,
receives a scalar reward $r(t)=R\big(s(t),a(t)\big)$, and transitions to the next state
$s(t+1)$ according to the transition kernel $P(\cdot\mid s(t),a(t))$.

As illustrated in Fig.~\ref{interaction auv env} we adopt a centralized PPO framework in which a single actor–critic architecture operates on the global state and outputs a joint action vector for all AUVs. The policy is trained using clipped surrogate objectives and generalized advantage estimation, enabling stable updates while capturing the coupling introduced by shared AoI dynamics, sensor energy evolution,and inter-AUV interactions.

\subsubsection{State Space}

For a system with $N$ AUVs and $K$ sensor nodes, the system state at time $t$ is defined as
\begin{equation}
\begin{aligned}
s(t)=\Big(
 \{\boldsymbol{\ell}_i(t),&\theta_i(t),v_i(t)\}_{i=1}^{N}, \;
\mathbf{A}(t),\;
\mathbf{e}(t), \\
&
\{\|\boldsymbol{\ell}_k-\boldsymbol{\ell}_i(t)\|\}_{\substack{i=1,\ldots,N \\ k=1,\ldots,K}}
\Big),
\end{aligned}
\label{eq:state_def_multi}
\end{equation}
where $\boldsymbol{\ell}_i(t)=[x_i(t),y_i(t)]^\top$ denotes the position of AUV $i$,
$\theta_i(t)$ and $v_i(t)$ are its heading and speed, respectively,
$\mathbf{A}(t)=[A_1(t),\ldots,A_K(t)]$ is the AoI vector of all sensor nodes, and
$\mathbf{e}(t)=[e_1(t),\ldots,e_K(t)]$ denotes the available energy at each node. Note that the relative position vectors and distances supply geometry-aware information essential for motion planning, distance-dependent acoustic propagation, and coordination. In the single-AUV case, terms for other AUVs are masked and excluded, while the remaining elements of \eqref{eq:state_def_multi} stay unchanged.

\subsubsection{Action Space}

The action space is defined as a joint discrete space over all AUVs. At time slot $t$,
the centralized policy selects a joint action
\begin{equation}
a(t)=\big(a_1(t),a_2(t),\ldots,a_N(t)\big),
\end{equation}
where $a_i(t)$ denotes the action assigned to AUV~$i$.

The action of AUV~$i$ is given by
\begin{equation}
a_i(t)=\big(
\tilde{\Delta\theta}_i(t),\;
\tilde{\Delta v}_i(t),\;
k^{\mathrm{WET}}_i(t),\;
k^{\mathrm{DATA}}_i(t)
\big),
\label{eq:action_def}
\end{equation}
where $\tilde{\Delta\theta}_i(t)\in\{0,\ldots,K_\theta-1\}$ and
$\tilde{\Delta v}_i(t)\in\{0,\ldots,K_v-1\}$ are discrete control indices for heading and
speed adjustments, respectively; $K_\theta$ and $K_v$ denote the numbers of
discretization levels for heading and speed commands.
These indices are mapped to bounded physical increments according to \vspace{-1.8mm}
\begin{align}
\Delta\theta_i(t) &=
\left(\frac{2\,\tilde{\Delta\theta}_i(t)}{K_\theta-1}-1\right)\Delta\theta_{\max}, \\
\Delta v_i(t) &=
\left(\frac{2\,\tilde{\Delta v}_i(t)}{K_v-1}-1\right)\Delta v_{\max},
\end{align}
so that the feasibility constraints on the incremental controls in~(12e) and (12f) are satisfied.

The variables $k^{\mathrm{WET}}_i(t)\in\{1,\ldots,K\}$ and
$k^{\mathrm{DATA}}_i(t)\in\{1,\ldots,K\}$ select the sensor nodes targeted by AUV~$i$
for acoustic wireless energy transfer and uplink data transmission, respectively. This
formulation enables each AUV to simultaneously decide its motion and communication actions at every time slot.

For each AUV, the number of admissible discrete action combinations is
$K_\theta K_v K^2$. Consequently, for a system with $N$ AUVs, the joint action space has
cardinality $(K_\theta K_v K^2)^N$, which grows exponentially with the number of AUVs.For a moderate number of AUVs and sensor nodes, this complexity remains tractable, and stable convergence is observed in practice using an efficient centralized PPO framework.

\subsubsection{Reward Function}

The reward function is designed to (i) drive the AUVs toward the docking zone,
(ii) reduce the AoI of all nodes, (iii) promote fairness in node servicing,
(iv) penalize infeasible or unproductive motion, (v) encourage energy-feasible
transmissions, and (vi) discourage unsafe proximity between AUVs.

Let $d_i(t)\triangleq \|\boldsymbol{\ell}_i(t)-\boldsymbol{\ell}_{\mathrm{g}}\|$ denote
the distance between AUV~$i$ and the docking center
$\boldsymbol{\ell}_{\mathrm{g}}$, and let
$\bar{A}(t)\triangleq \frac{1}{K}\sum_{k=1}^{K}A_k(t)$ be the average AoI. The Jain's
fairness index $\mathcal{J}(t)$ is computed from the empirical service counts as in
\eqref{eq:jain_env}. The instantaneous reward is expressed as \vspace{-0.5mm}
\begin{align}\label{eq:reward_env}
r(t)\!= {} &
\alpha_{\mathrm{g}}
\sum_{i=1}^{N}\big[d_i(t{-}1)-d_i(t)\big]
-\alpha_{\mathrm{a}}\,\bar{A}(t)
-\!\alpha_{\mathrm{f}}\big(1-\mathcal{J}(t)\big) \notag \\
& -\rho_{\mathrm{bd}}(t)
-\rho_{\mathrm{st}}(t)
+\rho_{\mathrm{m}}(t)
-\rho_{\mathrm{col}}(t)
\!+r_{\mathrm{dock}}(t),
\end{align} 
where $\alpha_{\mathrm{g}},\alpha_{\mathrm{a}},\alpha_{\mathrm{f}}>0$ are weighting
coefficients that were selected empirically to prioritize AoI reduction while treating fairness, collision avoidance, and energy-margin terms as regularization components. The same set of coefficients was used across all simulations.

The terms $\rho_{\mathrm{bd}}(t)$ and $\rho_{\mathrm{st}}(t)$ penalize boundary
violations and near-zero displacement (stalling), respectively, reflecting the
implementation in which out-of-bounds motion incurs a large penalty and negligible
movement incurs a smaller penalty. The term $r_{\mathrm{dock}}(t)$ provides a fixed
positive reward when an AUV first enters the docking zone, together with an additional
bonus when all AUVs have successfully docked.

Collision avoidance is enforced through a smooth distance-margin penalty. Let
$d_{ij}(t)=\|\boldsymbol{\ell}_i(t)-\boldsymbol{\ell}_j(t)\|$ denote the inter-AUV
distance. When at least one AUV is outside the docking zone, the collision penalty follows a Gaussian-distance based model given by 
\begin{equation}
\label{eq:collision_penalty}
\rho_{\mathrm{col}}(t)
\!=\!
-\alpha_{\mathrm{c}}
\sum_{i < j}
\left(
1 - \!
\exp\!\left(
-\frac{
\big[\max\!\big(0,\, d_{\mathrm{th}}\! - d_{ij}(t)\big)\big]^2
}{2\sigma_{\mathrm{c}}^2}
\right) \!
\right),
\end{equation}
where $d_{ij}(t)$ is the distance between the two AUVs, $d_{\mathrm{th}}$ is the minimum
safety distance, $\sigma_{\mathrm{c}}$ controls the smoothness of the penalty inside the
unsafe region, and $\alpha_{\mathrm{c}}$ sets the penalty scale. The term is inactive
when the separation exceeds $d_{\mathrm{th}}$ or both AUVs are inside the docking zone.

Finally, we include an energy-margin shaping term that promotes selecting nodes whose stored acoustic energy exceeds the required uplink transmission energy. For AUV~$i$, let $k^{\mathrm{DATA}}_i(t)$ denote the selected data node and $E_{\mathrm{req},k^{\mathrm{DATA}}_i(t)}(t)$ the required energy. The shaping term is
\begin{equation}
\label{eq:margin_reward}
\rho_{\mathrm{m}}(t)
=
\frac{\alpha_{\mathrm{m}}}{N}
\sum_{i=1}^{N}
\tanh\!\left(
\frac{e_{k^{\mathrm{DATA}}_i(t)}(t)
-
E_{\mathrm{req},k^{\mathrm{DATA}}_i(t)}(t)}
{E_{\mathrm{req},k^{\mathrm{DATA}}_i(t)}(t)+\varepsilon}
\right),
\end{equation}
where $\alpha_{\mathrm{m}}>0$ controls the shaping strength and $\varepsilon>0$ ensures numerical stability. This bounded margin term encourages energy-feasible transmissions while softly penalizing insufficient energy cases.

\begin{figure}[t!]
    \centering
    \begin{subfigure}{0.491\columnwidth}
        \centering
        \includegraphics[width=\linewidth]{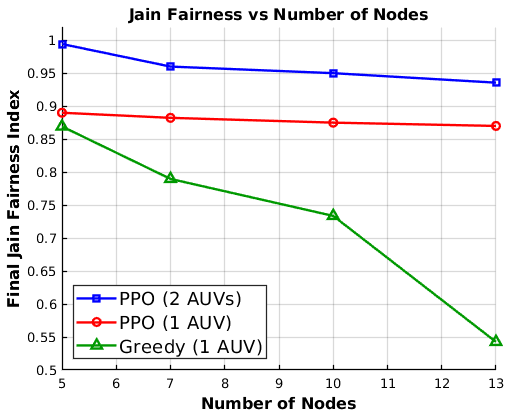}
        \caption{Fairness Index}
        \label{fig:fairness}
    \end{subfigure}
    \begin{subfigure}{0.491\columnwidth}
        \centering
        \includegraphics[width=\linewidth]{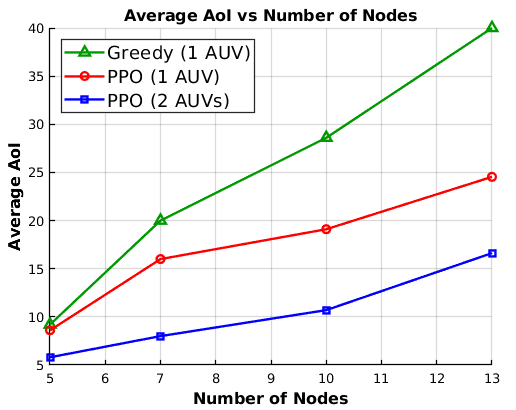}
        \caption{AoI}
        \label{fig:aoi}
    \end{subfigure}

    \vspace{-1mm}
    \caption{Performance comparison of the proposed and benchmark schemes for different network sizes. }
    \label{fig:fairness_aoi}
    \vspace{-4mm}
\end{figure}

\vspace{-4mm}
\section{Numerical Results}\label{sec:results} \vspace{-1mm}
\begin{figure*}[t!]
  \centering
  \subfloat[RL, 1 AUV\label{fig:traj_rl_1auv}]{
    \includegraphics[width=0.3\textwidth]{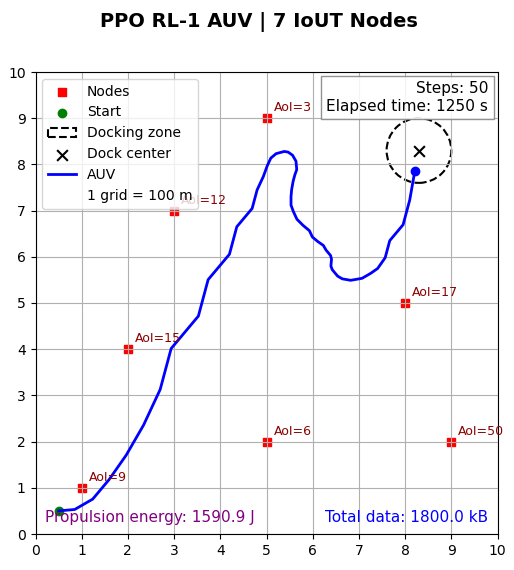}}
    \hspace{2mm}
  \subfloat[RL, 2 AUVs\label{fig:traj_rl_2auv}]{
    \includegraphics[width=0.3\textwidth]{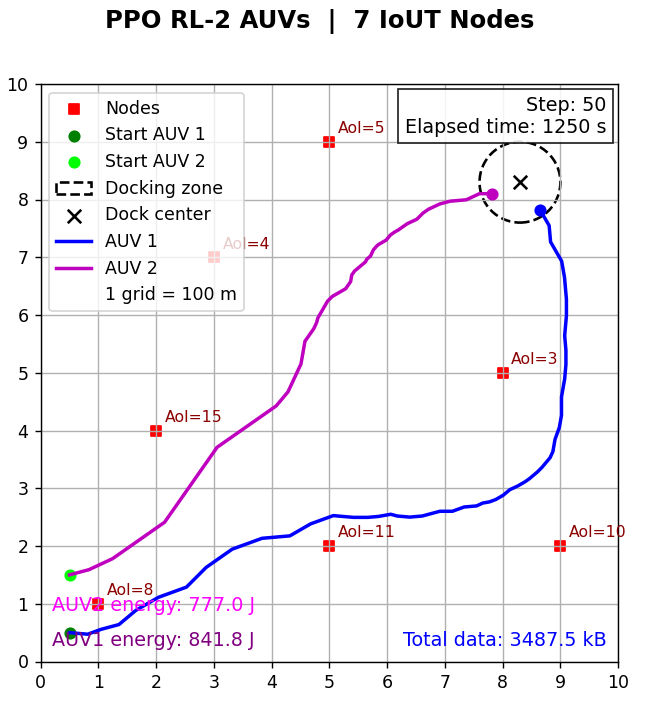}}
    \hspace{2mm}
  \subfloat[Greedy, 1 AUV\label{fig:traj_greedy}]{
    \includegraphics[width=0.3\textwidth]{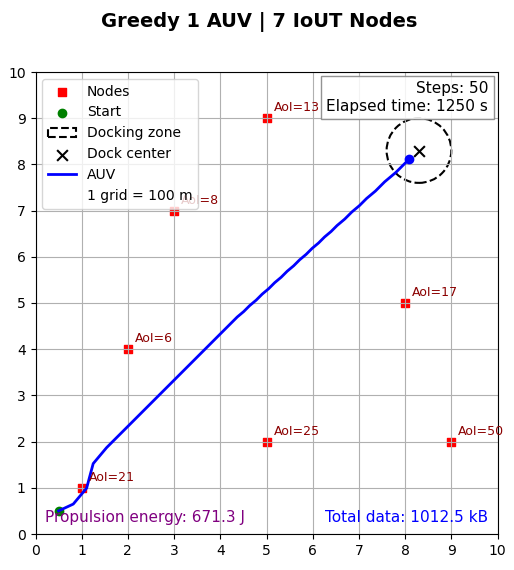}}

  \caption{AUV trajectories and total collected data for a network with 7 IoUT nodes under different scheduling strategies.}
  \label{fig:traj_triptych}
  \vspace{-2mm}
\end{figure*}
The simulation parameters used in this section are summarized in Tables~\ref{tab:motion} and~\ref{tab:acoustic}. While the proposed formulation applies to a general number of AUVs $N$, we focus on the case of two coordinated AUVs as a representative multi-agent setting to illustrate the benefits of cooperative trajectory planning compared to single-agent strategies. Fig.~\ref{fig:fairness_aoi} compares the average AoI and Jain fairness index of the proposed PPO-based RL schemes and the greedy benchmark under different network sizes. The RL approach with two AUVs consistently achieves the lowest average AoI while maintaining the highest fairness index, benefiting from spatial parallelism and balanced node servicing. In the single-AUV case, PPO also outperforms the greedy strategy in both AoI and fairness, and this advantage becomes more pronounced as the number of IoUT nodes increases.

This trend stems from the increasing complexity of the scheduling problem in larger networks. As the number of nodes grows, the AUV must distribute its limited service time among more devices, making the AoI evolution highly sensitive to revisit decisions. The greedy policy relies primarily on instantaneous geometric criteria and does not account for long-term AoI accumulation, which results in service imbalance when the network scales. In contrast, the PPO-based strategies learn coordinated revisit patterns that better regulate AoI growth, thereby preserving fairness and improving overall performance in larger systems.

\begin{figure}[t!]
    \centering
    \begin{subfigure}{0.491\columnwidth}
        \centering
        \includegraphics[width=\linewidth]{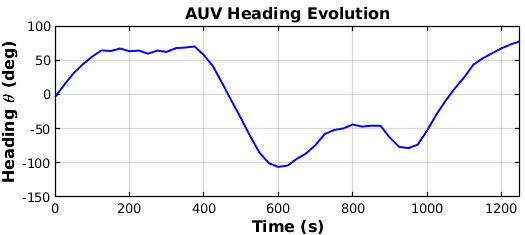}
        \caption{Heading Angle}
        \label{fig:heading}
    \end{subfigure}
    \begin{subfigure}{0.491\columnwidth}
        \centering
        \includegraphics[width=\linewidth]{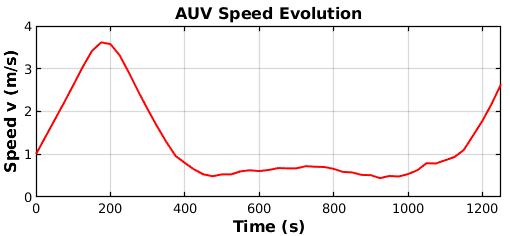}
        \caption{Speed}
        \label{fig:speed}
    \end{subfigure}

    \vspace{-1mm}
    \caption{AUV speed and heading evolution for a network with 7 IoUT nodes
underand PPO-based RL scheduling using a single AUV.}
    \label{fig:speed_heading}
    \vspace{-3mm}
\end{figure}

These trends are further illustrated in Fig.~\ref{fig:traj_triptych} for the 7-node scenario. In the single-AUV case, the PPO policy learns to steer the vehicle closer to multiple nodes before proceeding toward the docking region, resulting in more uniform information updates. This behavior is reflected in the smooth speed and heading evolution shown in Fig.~\ref{fig:speed_heading}, where the RL controller adjusts both velocity and orientation to adapt to node locations, unlike the greedy strategy which follows an almost fixed heading. For the two-AUV configuration, the learned policies implicitly partition the environment, with each AUV servicing a distinct spatial region, reducing revisit delays and improving fairness. Consequently, the RL schemes collect more data within the same time horizon. Although the greedy policy consumes less propulsion energy due to its minimal maneuvering, this comes at the expense of higher AoI, reduced fairness, and lower data collection.

\vspace{-1.2mm}
\begin{table}[t]
\centering
\scriptsize
\setlength{\tabcolsep}{3pt}

\begin{minipage}{0.48\columnwidth}
\centering
\caption{Motion \& Control}
\label{tab:motion}
\begin{tabular}{cc|cc}
\toprule
Param. & Val. & Param. & Val. \\
\midrule
$V_{\max}$ & 4 m/s & $C_d$ & 0.006 \\
$S$ & 3 m$^2$ & $\eta_{\text{prop}}$ & 0.7 \\
$\rho$ & 1000 kg/m$^3$ & $H$ & 40 W \\
$\Delta\theta_{\max}$ & 25$^\circ$ & $\Delta v_{\max}$ & 0.4 m/s \\
$t$ & 25 s & $T_{\max}$ & 55 slots \\
$d_{\text{th}}$ & 100 m & $d_{dock}$ & 0.7 \\
\bottomrule
\end{tabular}
\end{minipage}
\hfill
\begin{minipage}{0.48\columnwidth}
\centering
\caption{Acoustic \& Comm.}
\label{tab:acoustic}
\begin{tabular}{cc|cc}
\toprule
Param. & Val. & Param. & Val. \\
\midrule
$f_{\text{WET}}$ & 70 kHz & $f_{\text{data}}$ & 50 kHz \\
$P_{\text{tx}}$ & 5 W & $\eta_{\text{tx}}$ & 0.7 \\
$DI_{\text{tx,rx}}$ & 10 dB & $k_s$ & 1.5 \\
RVS & -150 dB & $R_p$ & 125 $\Omega$ \\
$n_{\text{hyd}}$ & 4 & $B$ & 1 kHz \\
$K_{\text{reset }}$ & 3 & $R$ & 12 kbps \\
\bottomrule
\end{tabular}
\end{minipage}

\vspace{-3mm}
\end{table}

\section{Conclusions} \label{sec:Conclusions} \vspace{-0.5 mm}
This letter presented a propulsion-aware DRL framework for coordinated multi-AUV trajectory control and acoustic energy-assisted data collection in IoUT networks. By integrating continuous kinematic control, energy feasibility, fairness regulation, and docking constraints within a centralized PPO formulation, the proposed approach enables scalable AoI-aware coordination among multiple AUVs. Numerical results demonstrate consistent improvements in information freshness, fairness, and data collection efficiency compared with structured heuristic baselines. Future work will extend the framework toward larger AUV fleets and hybrid underwater communication modalities.  \vspace{-1.8mm}

\bibliographystyle{IEEEtran}
\bibliography{references}
\end{document}